# Geometric Control of Pairing: Universal Scaling of Superconductivity at KTaO$_3$ Interfaces

Xueshan Cao[1,*], Meng Zhang[1,*,†], Yishuai Wang[1,*], Ming Qin[1], Yi Zhou[2,‡], and Yanwu Xie[1,3§]

[1]School of Physics, and State Key Laboratory for Extreme Photonics and Instrumentation, Zhejiang University, Hangzhou 310027, China

[2]Institute of Physics, Chinese Academy of Sciences, Beijing 100190, China

[3]Hefei National Laboratory, Hefei 230088, China.

[*]These authors contributed equally to this work.

[†]Contact author: physmzhang@zju.edu.cn

[‡]Contact author: yizhou@iphy.ac.cn

[§]Contact author: ywxie@zju.edu.cn

**Abstract.**

   The superconducting transition temperature $T_c$ at KTaO$_3$-based oxide interfaces exhibits a dramatic dependence on crystallographic orientation, yet a unifying principle has remained elusive. Here, we discover a universal linear scaling between $T_c$ and a single geometric parameter — the angle $\theta$ between the (*hkl*) plane and the (100) plane — across ten different orientations of LaAlO$_3$/KTaO$_3$ interfaces. With the exception of (100), all orientations exhibit two-dimensional superconductivity, with transition temperatures $T_c$ ranging from ∼ 0.12 K to 1.9 K. This linear $\theta - T_c$ scaling is robust against variations in growth temperature, device geometry, and transport configuration. By establishing geometric orientation as a direct control knob for pairing strength, our results impose a critical benchmark for microscopic theories of superconductivity in KTaO$_3$-based systems.

**Introduction** — Oxide interfaces are fertile grounds for emergent quantum phenomena, hosting two-dimensional (2D) superconductivity [1–6], magnetism [7–11], ferroelectricity [11–13], and nonreciprocal transport [14,15]. Among these, KTaO$_3$ -based interfaces have emerged as a premier platform, boasting superconducting transition temperatures $T_c$ up to ~2 K [4–6] — an order of magnitude higher than SrTiO$_3$-based counterparts. This enhancement is attributed to the Ta 5d orbitals, which impart strong spin-orbit coupling (SOC) that is further amplified by inversion-symmetry breaking [16–19].

A distinctive feature of KTaO$_3$ superconductivity is its extreme sensitivity to crystallographic orientation [4–6,20–24]: $T_c \approx 2$ K for (111), $\approx 0.9$ K for (110), and $\approx 0.25$ K for (100) (Recent studies [21] indicate that $T_c$ at CaZrO$_3$/ KTaO$_3$ (100) can reach 0.25 K, while the LaAlO$_3$/ KTaO$_3$ (100) interface remains normal down to 25 mK [22]). This strong anisotropy has been variously attributed to orientation-specific orbital structures [22] or anisotropic electron-phonon coupling [23]. However, prior studies have been restricted to principal planes, leaving it unclear whether these differences represent discrete electronic states or points on a continuous geometric trend.

In this work, we investigate LaAlO$_3$/KTaO$_3$ interfaces across ten distinct crystallographic orientations (Fig. 1 and Fig. S1). We demonstrate that $T_c$ is governed universally by the geometric angle $\theta$ between the $(hkl)$ plane and the (100) plane. This linear scaling persists regardless of growth conditions or sample patterning, suggesting that interface geometry acts as a continuous tuner of the pairing interaction.

We fabricated heterostructures on ten different KTaO$_3$ orientations (see Supplementary Material (SM) Note 1 for detailed methods), ranging from principal planes to high-index surfaces [Fig. 1(b)]. KTaO$_3$ substrates of all orientations exhibit atomically flat surfaces, which are maintained after LaAlO$_3$ film deposition (Fig. S2 and S3). X-ray diffraction (XRD) indicates that the LaAlO$_3$ films are amorphous [Fig. S4(a)], with a uniform thickness of 10 nm across all orientations [Fig. S4(b)]. Detailed transport measurements using Hall-bar geometries [Fig. 2(a)] reveal a clear superconducting transition for all orientations except (100). The transitions are sharp [Figs. 2(b),(c) and Fig. S5], and voltage-current $(V - I)$ characteristics (Fig. S6) follow a $V \propto I^3$ power law consistent with Berezinskii–Kosterlitz–Thouless (BKT) physics [6]. The BKT transition temperatures align with the resistance midpoints [Figs. S6(b),(f)], and strong anisotropy in the upper critical fields ($H_{c2}$) (detailed below) confirms the 2D nature of the superconducting state across all superconducting samples.

To identify the driver of the variations in $T_c$, we analyzed crystallographic parameters. As shown in Figs. 3(a) and 3(b), $T_c$ shows no correlation with interplanar spacing $d_{(hkl)} = \frac{a}{\sqrt{h^2+k^2+l^2}}$, where $a$ is the cubic unit cell parameter. However, a striking unification emerges when $T_c$ is plotted against $\theta = \arccos \frac{h}{\sqrt{h^2+k^2+l^2}}$, the angle of the surface normal relative to the (100) axis, where $l \leq k \leq h$. As displayed in Fig. 3(c), $T_c$ values for all nine superconducting orientations collapse onto a single linear trajectory: $T_c \propto \theta$. This scaling holds for samples grown at both 300 °C and 680 °C, and for both patterned and unpatterned (bare) films, indicating that $\theta$ is an intrinsic descriptor of the superconducting ground state.

This geometric universality provides immediate constraints on the pairing mechanism. Unlike standard BCS paradigms where $T_c$ is often driven by the density of states, we find that $T_c$ is largely decoupled from the Hall carrier density $n_s$ and the mobility $\mu$. As shown in Fig. 4(a), where all the samples grown at $T_g$ = 300 °C, excluding the non-superconducting (100) case, $n_s$ remains nearly constant (6.3–7.1 × $10^{13}$ cm$^{-2}$) across orientations despite $T_c$ varying by an order of magnitude. Similar phenomena were observed at $T_g$ = 680 °C (Fig. S7). This reinforces observations in (111) and (110) interfaces [5,22,25] that KTaO$_3$ superconductivity is distinct from the dome-like doping dependence of SrTiO$_3$.

Instead, the geometric angle $\theta$ appears to regulate the spatial confinement of the superconducting condensate. By analyzing the temperature dependence of anisotropic upper critical fields (Figs. S8, S9), we extracted the superconducting layer thickness $d_{sc}$ and coherence lengths $\xi_{GL}$ using Ginzburg-Landau theory [6,26]. We observe an inverse relationship: $T_c$ increases as the superconducting layer thickness $d_{sc}$ decreases [Fig. 4(b)]. Since the sheet carrier density $n_s$ is nearly constant, a thinner $d_{sc}$ implies a higher effective 3D carrier density ($n_{3D} = n_s/d_{SC}$), leading to an increase in $T_c$ with $n_{3D}$ (Fig. S10).

Furthermore, the confinement is linked to spin-orbit interactions. The Rashba SOC splitting energy $\Delta_{SO}$, estimated from the upper critical fields (see SM note 2 for details) [25,27], displays positive correlation with $T_c$ across the investigated orientations [Fig. 4(c)]. This suggests that increasing the angle $\theta$ [moving away from (100)] enhances the structural inversion asymmetry or orbital overlap in a way that simultaneously tightens confinement, boosts SOC, and strengthens pairing.

**Discussions** — The extreme orientation dependence of $T_c$ in LaAlO$_3$/KTaO$_3$ interfaces has previously been interpreted within two main frameworks. The first invokes orientation-specific orbital configurations in the interfacial quantum well, which modify the density of states and pairing interaction [22]. The second focuses on anisotropic electron–phonon coupling to surface phonon modes, supported by photoemission evidence for orientation-dependent phonon spectra in KTaO$_3$ [23]. Both mechanisms are plausible in light of strong inversion-symmetry breaking and large Ta-derived spin-orbit coupling [16,17], but they have been tested only against data from the principal (100), (110), and (111) planes.

Our expanded dataset demonstrates that $T_c$ values for seven additional superconducting orientations, grown under two distinct conditions, collapse onto a single linear function of $\theta$ — the geometric angle between the (*hkl*) plane and the (100) plane. This universality persists regardless of carrier density, mobility, thickness variation, or device geometry, pointing to $\theta$ as an intrinsic descriptor of the pairing environment. The observed scaling may reflect a continuous evolution of orbital overlap, phonon mode character, or Rashba SOC strength with $\theta$, providing a coarse-grained geometrical parameter that integrates multiple microscopic effects.

From an applied perspective, the $\theta - T_c$ relation offers a deterministic route to controlling pairing strength via substrate orientation. Orientation engineering can be combined with established tuning methods such as electrostatic gating [5,25], strain [28], or heterostructure design to maximize $T_c$ for device applications. Our results therefore both narrow the landscape of plausible pairing mechanisms — whether orbital, phononic, SOC-driven, or a combination thereof — and establish geometric orientation as a central design principle for KTaO$_3$-based superconductors.

**Conclusion** —The discovery of linear $\theta - T_c$ scaling fundamentally reframes the understanding of KTaO$_3$ interfaces. Previous models relying on discrete orbital symmetries of principal planes must now be generalized to account for this continuous geometric evolution. Whether the mechanism is driven by anisotropic electron-phonon coupling [23] or orbital mixing in the quantum well [22], a successful theory must reproduce this linear dependence on $\theta$.

Our findings establish geometric orientation as a precise, deterministic tool for engineering oxide superconductivity. By selecting specific crystal cuts, one can continuously tune the pairing strength, SOC, and confinement width. This geometric "knob", combined with electrostatic gating, opens new pathways for designing higher-$T_c$ oxide devices.

*Acknowledgments*—We thank Prof. Chao Cao for fruitful discussions. Y.X. acknowledges support from the National Natural Science Foundation of China (NSFC) (Grant Nos. 12325402 and 12534005), the National Key R&D Program of China (Grant No. 2023YFA1406400), and the Quantum Science and Technology-National Science and Technology Major Project (Grant No. 2021ZD0300200). M.Z. is supported by the NSFC (Grant No. 12504226). Y.Z. is supported by the National Key R&D Program of China (Grant No. 2022YFA1403403) and the NSFC (Grant Nos. 12274441 and 12534004).

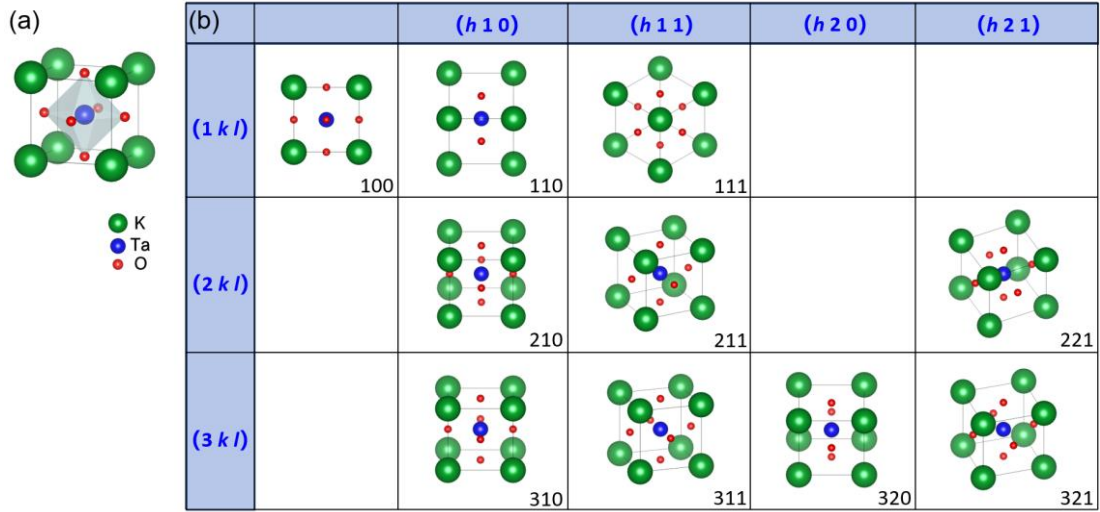

FIG. 1. Crystal structure and definition of geometric orientation. **(a)** Cubic perovskite unit cell of KTaO$_3$ (K: green, Ta: blue, O: red). **(b)** Top-view projections of the ten investigated ($hkl$) planes, ranging from the principal (100), (110), and (111) surfaces to high-index orientations. The central geometric parameter $\theta$ is defined as the angle between the surface normal of the ($hkl$) plane and the (100) axis. All samples are capped with a uniform 10-nm amorphous LaAlO$_3$ overlayer.

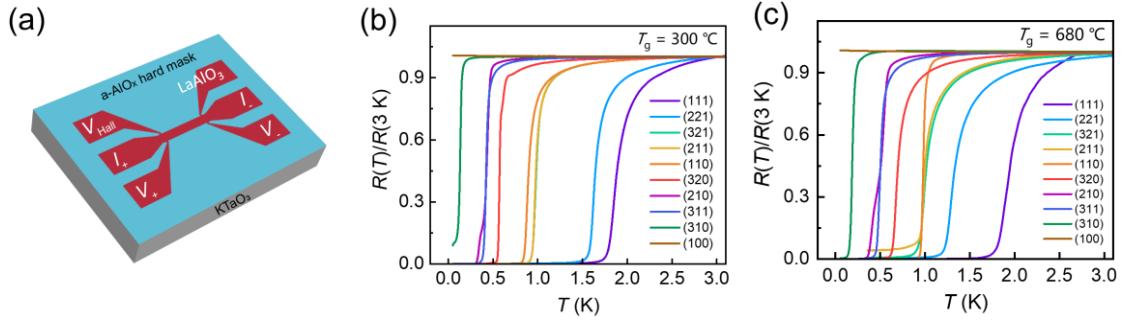

FIG. 2. Orientation-dependent superconducting transitions. **(a)** Schematic of the Hall-bar device geometry (channel dimension: $100 \times 20\ \mu m^2$) used for transport measurements. **(b)** Temperature-dependent sheet resistance $R(T)$, normalized to the normal-state resistance $R(3\text{K})$, for interfaces grown at $T_g = 300\,°C$. **(c)** Normalized resistance curves for interfaces grown at $T_g = 680\,°C$. With the exception of the (100) interface down to 50 mK, which remains metallic, all orientations exhibit sharp superconducting transitions.

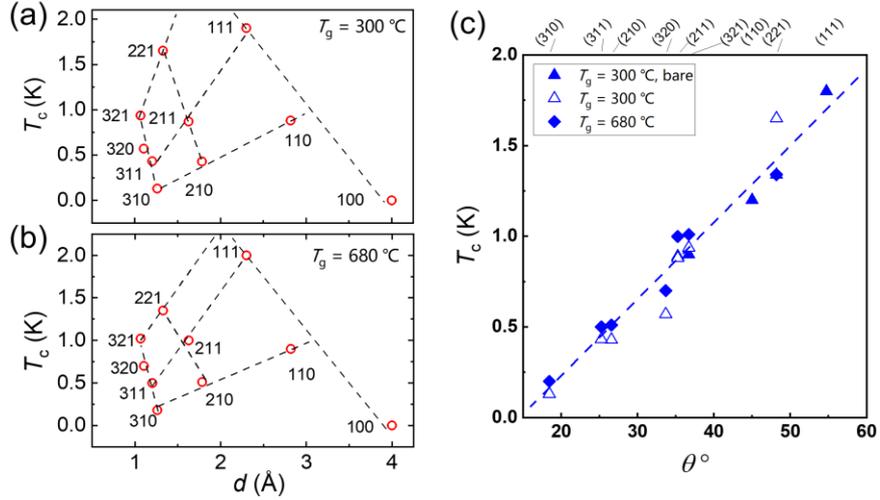

FIG. 3. Universal geometric scaling of $T_c$. **(a)** and **(b)** Superconducting transition temperature $T_c$ plotted against interplanar spacing $d_{(hkl)}$ for samples grown at **(a)** 300 °C and **(b)** 680 °C, respectively. No clear correlation is observed. **(c)** $T_c$ plotted against the geometric angle $\theta$. Data from all superconducting orientations, encompassing two growth temperatures and both patterned (Hall bar) and unpatterned ("bare") films, collapse onto a single linear trend. This indicates that $\theta$ is the primary descriptor of the pairing strength. Dashed lines are linear fits.

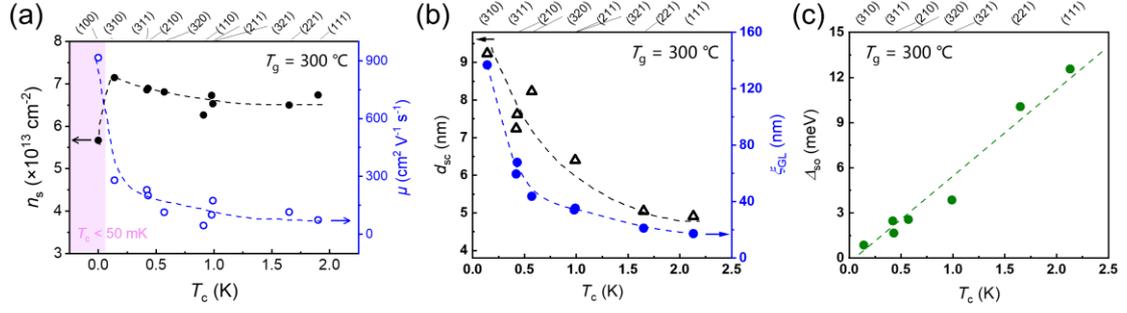

FIG. 4. Correlations between $T_c$ and microscopic parameters ($T_g$ = 300 °C). **(a)** Hall carrier density $n_s$ (left axis) and mobility $\mu$ (right axis) versus $T_c$. The pink shaded region indicates the non-superconducting (100) sample. For all superconducting orientations, $n_s$ is nearly constant ($\approx$ 6.3–7.1 × $10^{13}$ cm$^{-2}$), indicating $n_s$ is not driven by doping variations. **(b)** Superconducting layer thickness $d_{sc}$ and Ginzburg-Landau coherence length $\xi_{GL}(0)$ versus $T_c$. $T_c$ increases as the superconducting condensate becomes more spatially confined (smaller $d_{sc}$). **(c)** Rashba SOC splitting energy $\Delta_{SO}$ versus $T_c$, showing a positive correlation between SOC strength and the transition temperature.